\documentclass[apj]{emulateapj} 

\usepackage{graphicx}
\usepackage{ulem}
\usepackage{color}
\usepackage{amsmath}
\usepackage{amssymb}
\usepackage{float}
\usepackage{multirow}
\usepackage[stable]{footmisc}
\usepackage{subfigure}
\usepackage{setspace}
\usepackage{microtype}

\newcommand{\loggf}{\log gf}
\newcommand{\logZ}{\log Z/Z_\odot}
\newcommand{\FeH}{{\rm [Fe/H]}}
\newcommand{\epsFe}{\epsilon_{\rm Fe}}
\newcommand{\Teff}{T_{\rm eff}}
\newcommand{\Lc}{\lambda_{\rm c}}
\newcommand{\fsyn}{f_{\rm syn}}
\newcommand{\fsynr}{f_{\rm syn}^\dagger}

\newcommand{\kms}{{\rm km~s^{-1}}}

\shorttitle{\ion{Fe}{1} lines in 0.91--1.33~{$\mu$}m spectra of red giants}
\shortauthors{Kondo {et~al.}}

\begin{document}

\title{\ion{F\lowercase{e}}{1} lines in 0.91--1.33~{$\mu$}\lowercase{m} spectra of red giants\\
for measuring the microturbulence and metallicities}


\author{
Sohei~Kondo\altaffilmark{1},
Kei~Fukue\altaffilmark{1},
Noriyuki~Matsunaga\altaffilmark{2,1},
Yuji~Ikeda\altaffilmark{1,3},
Daisuke~Taniguchi\altaffilmark{2},
Naoto~Kobayashi\altaffilmark{1,4,5},
Hiroaki~Sameshima\altaffilmark{1},
Satoshi~Hamano\altaffilmark{1},
Akira~Arai\altaffilmark{1},
Hideyo~Kawakita\altaffilmark{1,6},
Chikako~Yasui\altaffilmark{1,7},
Natsuko~Izumi\altaffilmark{7},
Misaki~Mizumoto\altaffilmark{8},
Shogo~Otsubo\altaffilmark{6}, 
Keiichi~Takenaka\altaffilmark{6},
Ayaka~Watase\altaffilmark{6},
Akira~Asano\altaffilmark{6},
Tomohiro~Yoshikawa\altaffilmark{9},
and
Takuji~Tsujimoto\altaffilmark{7}
}

\altaffiltext{1}{Laboratory of Infrared High-resolution spectroscopy
(LiH), Koyama Astronomical Observatory, Kyoto Sangyo University,
Motoyama, Kamigamo, Kita-ku, Kyoto 603-8555, Japan:kondosh@cc.kyoto-su.ac.jp}
\altaffiltext{2}{Department of Astronomy, Graduate School of Science, The University of Tokyo, 7-3-1 Hongo, Bunkyo-ku, Tokyo 113-0033, Japan}
\altaffiltext{3}{Photocoding, 460-102 Iwakura-Nakamachi, Sakyo-Ku, Kyoto 606-0025, Japan}
\altaffiltext{4}{Kiso Observatory, Institute of Astronomy, School of
Science, the University of Tokyo, 10762-30 Mitake, Kiso-machi, Kiso-gun, Nagano 397-0101, Japan}
\altaffiltext{5}{Institute of Astronomy, Graduate School of Science, the
University of Tokyo, 2-21-1 Osawa, Mitaka, Tokyo 181-0015, Japan}
\altaffiltext{6}{Department of Physics, Faculty of Science, Kyoto Sangyo
University, Motoyama, Kamigamo, Kita-ku, Kyoto 603-8555, Japan}
\altaffiltext{7}{National Astronomical Observatory of Japan, 2-21-1 Osawa, Mitaka, Tokyo 181-0015, Japan}
\altaffiltext{8}{Centre for Extragalactic Astronomy, Department of Physics, University of Durham, South Road, Durham DH1 3LE, UK}
\altaffiltext{9}{Edechs, 17-203 Iwakura-Minami-Osagi-cho, Sakyo-ku, Kyoto 606-0003, Japan}


\begin{abstract}
For a detailed analysis of stellar chemical abundances, 
high-resolution spectra in the optical have mainly been used,
while the development of
near-infrared (NIR) spectrograph has opened new wavelength windows.
 Red giants have a large number of resolved absorption lines 
in both the optical and NIR wavelengths,
but the characteristics of the lines in different wave passbands are not necessarily the same.
We present a selection of \ion{Fe}{1} lines
in the $z^{\prime}$, $Y$, and $J$ bands (0.91--1.33~{$\mu$}m).
On the basis of two different lists of lines in this range,
the Vienna Atomic Line Database (VALD)
and the catalog published by Mel{\'e}ndez \& Barbuy in 1999 (MB99),
 we selected sufficiently strong lines that are not severely blended
and compiled lists with 107 \ion{Fe}{1} lines in total
(97 and 75 lines from VALD and MB99, respectively).
Combining our lists with high-resolution
($\lambda/\Delta\lambda = 28,000$) and high signal-to-noise ($>500$)
spectra taken with a NIR spectrograph, WINERED, 
we present measurements of
the iron abundances of two prototype red giants: Arcturus and $\mu$~Leo.
 A bootstrap method for determining the microturbulence
 and abundance together with their errors is
 demonstrated. The standard deviations of $\log\epsFe$ values from
 individual \ion{Fe}{1} lines
are significantly smaller when we use the lines from MB99 instead of
 those from VALD.
With the MB99 list, we obtained $\xi=1.20\pm0.11\ \kms$ and
$\log\epsFe=7.01\pm0.05$ dex for Arcturus, and $\xi=1.54\pm0.17\ \kms$
and $\log\epsFe=7.73\pm0.07$ dex for $\mu$~Leo. These final values show
better agreements with previous values in the literature
than the corresponding values we obtained with VALD.
\end{abstract}


\section{Introduction}

Recent developments in instruments (e.g.,\ multi-object spectrographs)
and statistical approaches (e.g., CANNON, \citealt{Ness-2015}; ASPCAP,
\citealt{GarciaPerez-2016}) provide opportunities to measure
metallicities of a larger number of stars and/or to higher precision.
Among the various methods available for estimating stellar
metallicities, the measurement of individual metallic lines in
high-resolution spectra is the most direct and fundamental one. Such
detailed analyses of high-resolution spectra have mostly been performed
with optical spectra, while recently developed instruments now
produce near-infrared (NIR) high-resolution spectra that are similarly
useful and high in quality. For example, the APOGEE project
established fiber-fed multi-object spectrographs to collect hundreds of
$H$-band spectra (1.5--1.7~{$\mu$}m, $\lambda/\Delta\lambda$ = 22,500)
simultaneously \citep{Majewski-2017}. Several other NIR spectrographs
with a single slit have been used for abundance analysis for individual
stars, especially those affected by strong interstellar extinction.
Such pioneering works include studies of chemical abundances of stars in
the Galactic bulge
\citep{Carr-2000,Cunha-2006,Ryde-2009,Ryde-2010,Ryde-2016} and red
supergiants in clusters in the inner disk
\citep{Davies-2009a,Davies-2009b,Origlia-2013,Origlia-2016}.

Since abundance analyses based on NIR spectra have now turned state
 of the art, they require, e.g., accurate calibration of oscillator strengths
 of absorption lines in that spectral domain.
For example, the APOGEE
project has not only measured the abundances of a large number of stars
but has also made progress in establishing methodology and fundamental
datasets: a list of absorption lines in the $H$ band
\citep{Shetrone-2015}, a new grid of atmospheric models
\citep{Meszaros-2012}, a tool to search for the best sets of stellar
parameters \citep{GarciaPerez-2016}, and so on. In particular, an
accurate line list is essential to perform chemical analysis in stellar
atmospheres. The correct identification of lines is mandatory, and
estimates of abundances cannot be accurately carried out without
accurate oscillator strengths. As \citet{Ryde-2009} pointed out, many lines in the
NIR are not properly identified or lack well-calibrated oscillator
strengths. Available line lists with a wide wavelength coverage include
Kurucz's database \citep{Kurucz-1995}, Vienna Atomic Line Database
\citep[VALD3;][]{Ryabchikova-2015}, and the list published by
\citeauthor{Melendez-1999} (1999; hereinafter referred to as MB99).
MB99 compiled absorption lines, which they identified in the solar
spectrum, and obtained astrophysical $\loggf$ values\footnote{Here and
elsewhere in this paper, we consider the logarithm to base 10.}. In
contrast, Kurucz's database and VALD3 have a significantly larger number
of lines including those only theoretically predicted. In this work, we
compared results of abundance analysis obtained with lines in the range
of 0.91--1.33~{$\mu$}m from VALD3 and the MB99 list, and also compared
our measurements with previous results.

In addition to comparing line lists, another goal of this study is to
test the determination of the microturbulence, $\xi$, using NIR
high-resolution spectra. In an abundance analysis of stars, $\xi$ is one
of the most important parameters, and its uncertainty often remains a
major error source for the metallicity. In a classical analysis of
optical high-resolution spectra, $\xi$ is estimated by necessitating
that $\log\epsFe$, defined as $\log(N_{\rm Fe}/N_{\rm H})+12$, from
individual lines shows no dependency on line strengths, e.g., equivalent
widths (EWs, denoted as $W$) or reduced EWs ($W/\lambda$). This method
requires a large number of iron lines with various strengths. For NIR
spectra, different methods for estimating $\xi$ have often been used so
far. \citet{Davies-2009b}, for example, obtained it by comparing the
molecular bands in synthetic and observed spectra. Sometimes $\xi$ is
assumed {\it a priori}. In an analysis of more than $10^5$ stars in the
APOGEE project, $\xi$ for giants were estimated from the relationship
between the surface gravity and $\xi$ in DR13 and by comparing observed
spectra to libraries of theoretical spectra in DR14
\citep{Holtzman-2018}\footnote{http://www.sdss.org/dr14/irspec/}.
In contrast, \citet{Smith-2013} estimated $\xi$
with $H$-band spectra in the same manner as the classical method
 mentioned above. However, the number of iron lines used was small
(eight or nine), and the range of their strengths was limited.
As shown below,
we can identify more lines with a broad range of strengths at 0.91--1.33~{$\mu$}m.

\section{Observation and Data Reduction}
We investigated WINERED spectra of well-studied red giants, Arcturus and
$\mu$~Leo. The former has a subsolar metallicity, and the latter is
significantly metal-rich; previous estimates are summarized in
Section~\ref{sec:atmosphere}. WINERED has a spectral resolution of
$R\equiv \lambda/\Delta\lambda \sim 28,000$. A single exposure covers a
wide wavelength range of 0.91--1.35~{$\mu$}m, which includes the
$z^{\prime}$, $Y$, and $J$ bands \citep{Ikeda-2016}. Such a wide
coverage is a huge advantage in abundance analysis. A large number of
\ion{Fe}{1} lines are included, and their strengths range from a
severely saturated regime to a very weak regime, thus allowing accurate
estimates of $\xi$.

We observed Arcturus and $\mu$~Leo on February 23, 2013 with WINERED
mounted on the Nasmyth focus of the 1.3~m Araki Telescope at Koyama
Astronomical Observatory, Kyoto Sangyo University, Japan
(Table~\ref{tab1}). WINERED is a cross-dispersed-type echelle
spectrograph using a 1.7~{$\mu$}m cutoff $2048 \times 2048$ HAWAII-2RG
array. The pixel scale is 0\farcs8~pixel$^{-1}$, and we used a slit
$48\arcsec$ in length and 1{\farcs}6 in width, providing a spectral
resolution of $R \sim 28,000$ \citep[further technical details are
described in][]{Ikeda-2016}. We also observed HIP~76267 (A1IV) as a
telluric standard. The total exposure times were 20, 240, and 600~s for
Arcturus, $\mu$~Leo, and HIP~76267, respectively. For every object, sky
frames without the target or any other visible stars included in the
slit were obtained to subtract the background including bias and dark of
the detector as well as the sky and ambient radiation.

\begin{deluxetable}{ccc}
\tabletypesize{\small}
\tablecaption{Targets and WINERED observations\label{tab1}}
\tablewidth{75mm}
\tablehead{
\colhead{}
& \colhead{Arcturus}
& \colhead{$\mu$~Leo}
}
\startdata
Alias & HD~124897, $\alpha$~Boo & HD~85503 \\ 
$T_{\rm eff}$~(K)$^\dagger$ & $4286 \pm 35$ & $4474 \pm 60$ \\ 
$\log g$~(dex)$^\dagger$ & $1.64 \pm 0.06$ & $2.51 \pm 0.09$ \\ 
$[{\rm M/H}$]~(dex)$^\dagger$ & $-0.52 \pm 0.08$ & $0.25 \pm 0.15$ \\ 
Date (UT) & 2013 Feb 23 & 2013 Feb 23 \\ 
Time (UT) & 16:23 & 17:18 \\ 
Exposures (s) & 20 (2~s$\times$ 10) & 240 (20~s$\times$ 12) \\ 
S/N$^{\ddagger}$ & 1200 & 900 \\ 
S/N$^{\ddagger\star}$ & 850 & 720 
\enddata
\tablenotetext{$\dagger$}{The stellar parameters are adopted from \citet{Heiter-2015}. For [M/H], we simply use their [Fe/H].}
\tablenotetext{$\ddagger$}{S/N is measured at around the middle of $J$ band. Note that these S/N consider statistical errors measured by comparing multiple integrations.}
\tablenotetext{$\star$}{After the correction of telluric lines with a spectrum (S/N$=$1200) of the telluric standard HIP~76267.}
\end{deluxetable}

All the data were reduced following standard procedures adopted in the
WINERED pipeline (Hamano et al., in preparation) that is established
using PyRAF,\footnote{PyRAF is a product of the Space Telescope Science
Institute, which is operated by AURA for NASA.} which calls IRAF
tasks,\footnote{IRAF is distributed by the National Optical Astronomy
Observatories, which are operated by the Association of Universities for
Research in Astronomy, Inc., under cooperative agreement with the
National Science Foundation.} including sky subtraction, scattered
light subtraction, flat-fielding (using a halogen lamp with an
integrating sphere), geometric transformation, aperture extraction, and
wavelength calibration based on Th-Ar lamp spectra. The continuum was
traced in each echelle order and normalized to unity. After the
pipeline reduction, we applied the method described in
\citet{Sameshima-2018} for the telluric correction. The spectrum of a
telluric standard, HIP~76267, with a high signal-to-noise ratio (S/N
{$\sim$}1200), was used for both Arcturus and $\mu$~Leo. The spectra in
different echelle orders were then combined by taking the averages at
overlapping wavelengths, and thus we obtained the continuum-normalized
continuous spectra of Arcturus and $\mu$~Leo for the $z^\prime$, $Y$,
and $J$ bands. The wavelength ranges of the three bands, in which the
telluric lines can be well corrected, cover 0.91--0.93, 0.96--1.115, and
1.16--1.33~{$\mu$}m, respectively. Finally, the stellar redshifts were
corrected so that the absorption lines can be directly compared with
those in synthetic spectra in rest air wavelength. We estimated the S/N
ratios at around 12,500~\AA, as given in Table~\ref{tab1}, in a manner
similar to that described in \citet{Fukue-2015}. Considering the noise
present in the telluric correction, we also calculated the S/N of the
spectra after the correction. The reduced spectra of Arcturus and
$\mu$~Leo are presented in Figure~\ref{fig1}.

\begin{figure*}
\begin{center}
\includegraphics[bb=80 80 640 640,width=0.98\hsize]{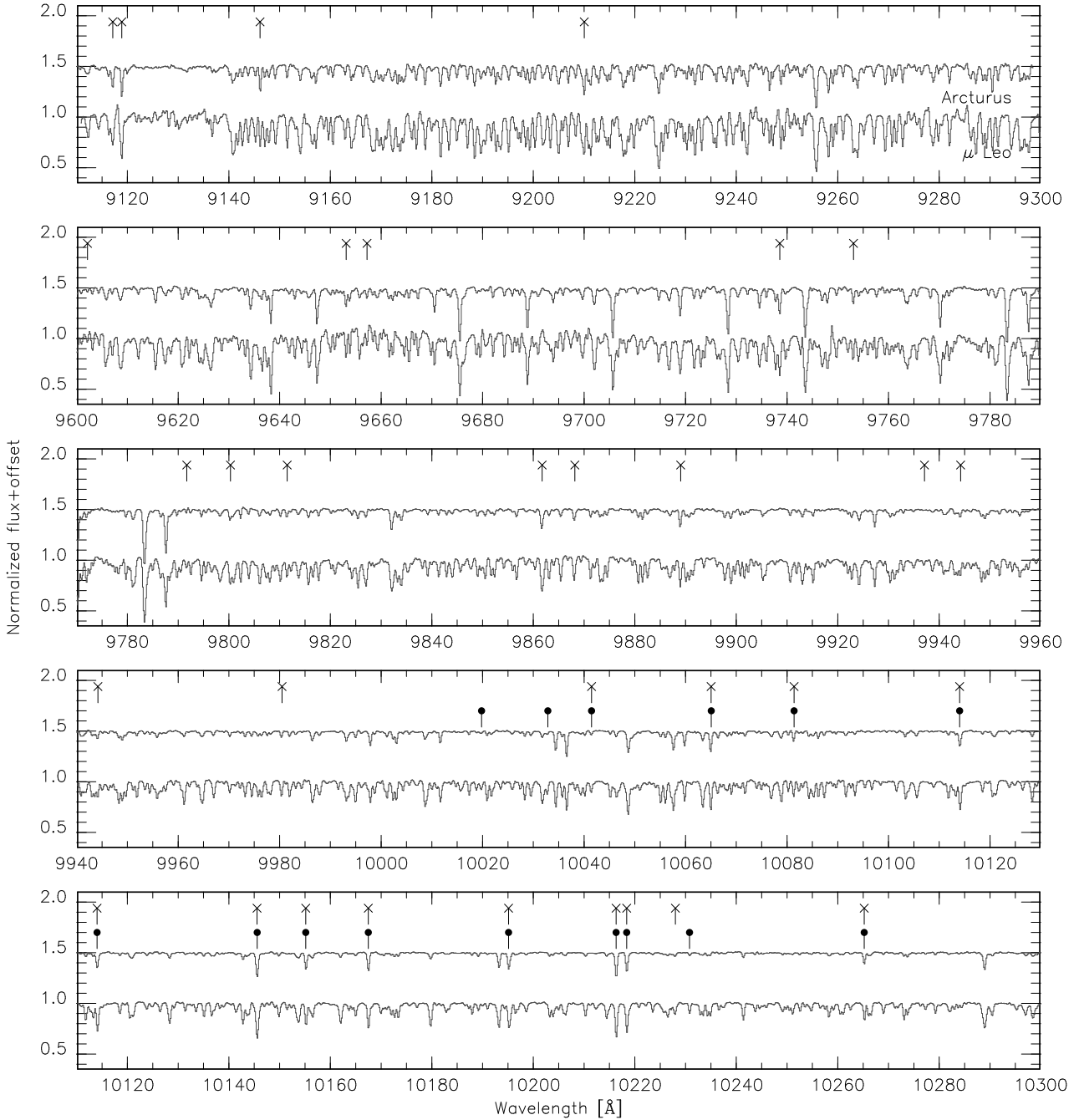}
\end{center}
\caption{Reduced spectra of Arcturus and $\mu$~Leo after the telluric
 correction. The short lines with the cross symbols and the filled circles indicate
\ion{Fe}{1} lines selected from VALD3 and MB99, respectively, for
the metallicity measurements for Arcturus and/or $\mu$~Leo.
\label{fig1}}
\end{figure*}

\addtocounter{figure}{-1}
\begin{figure*}
\begin{center}
\includegraphics[bb=80 80 640 640,width=0.98\hsize]{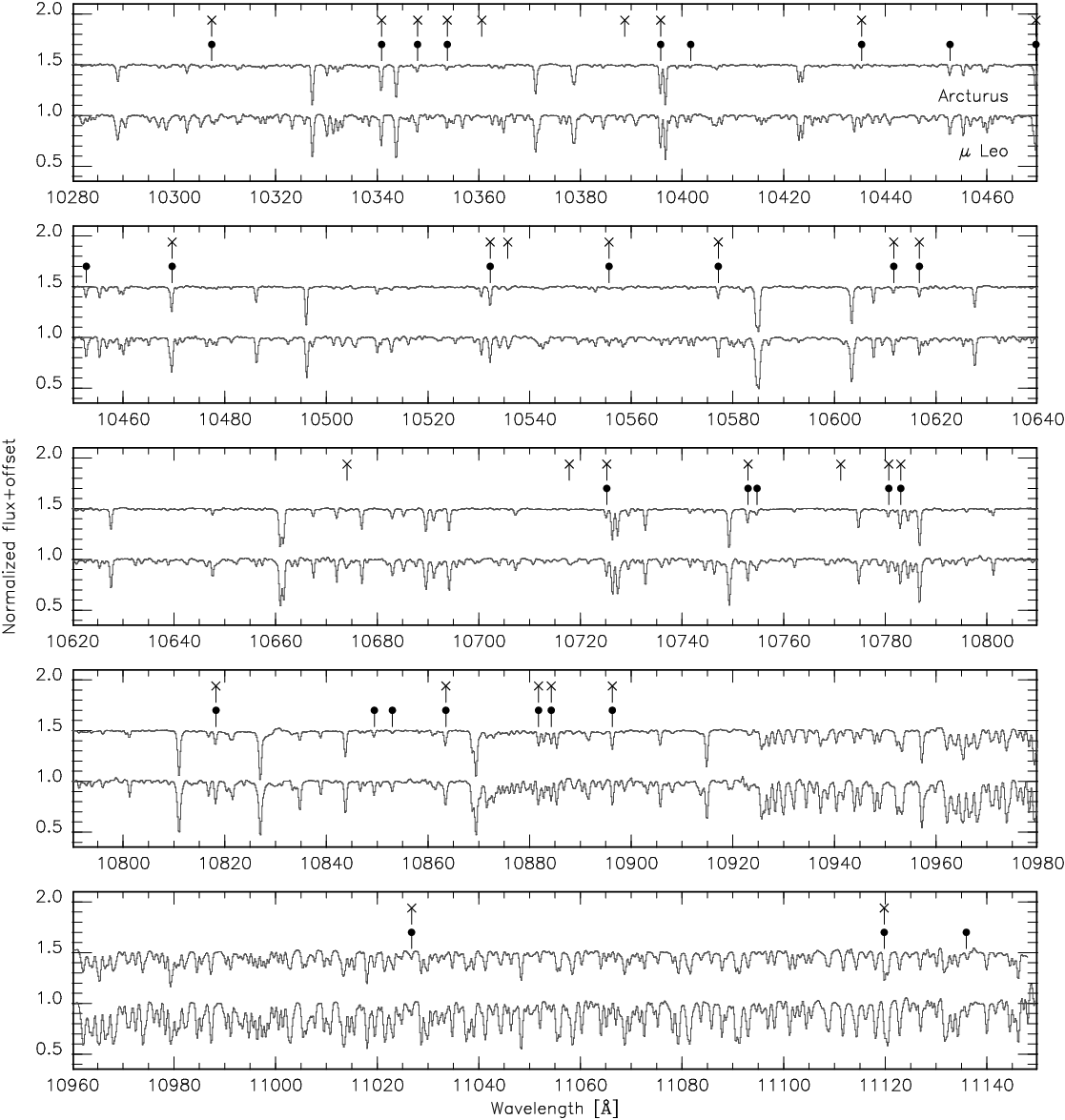}
\caption{continued.}
\end{center}
\end{figure*}
\addtocounter{figure}{-1}
\begin{figure*}
\begin{center}
\includegraphics[bb=80 80 640 640,width=0.98\hsize]{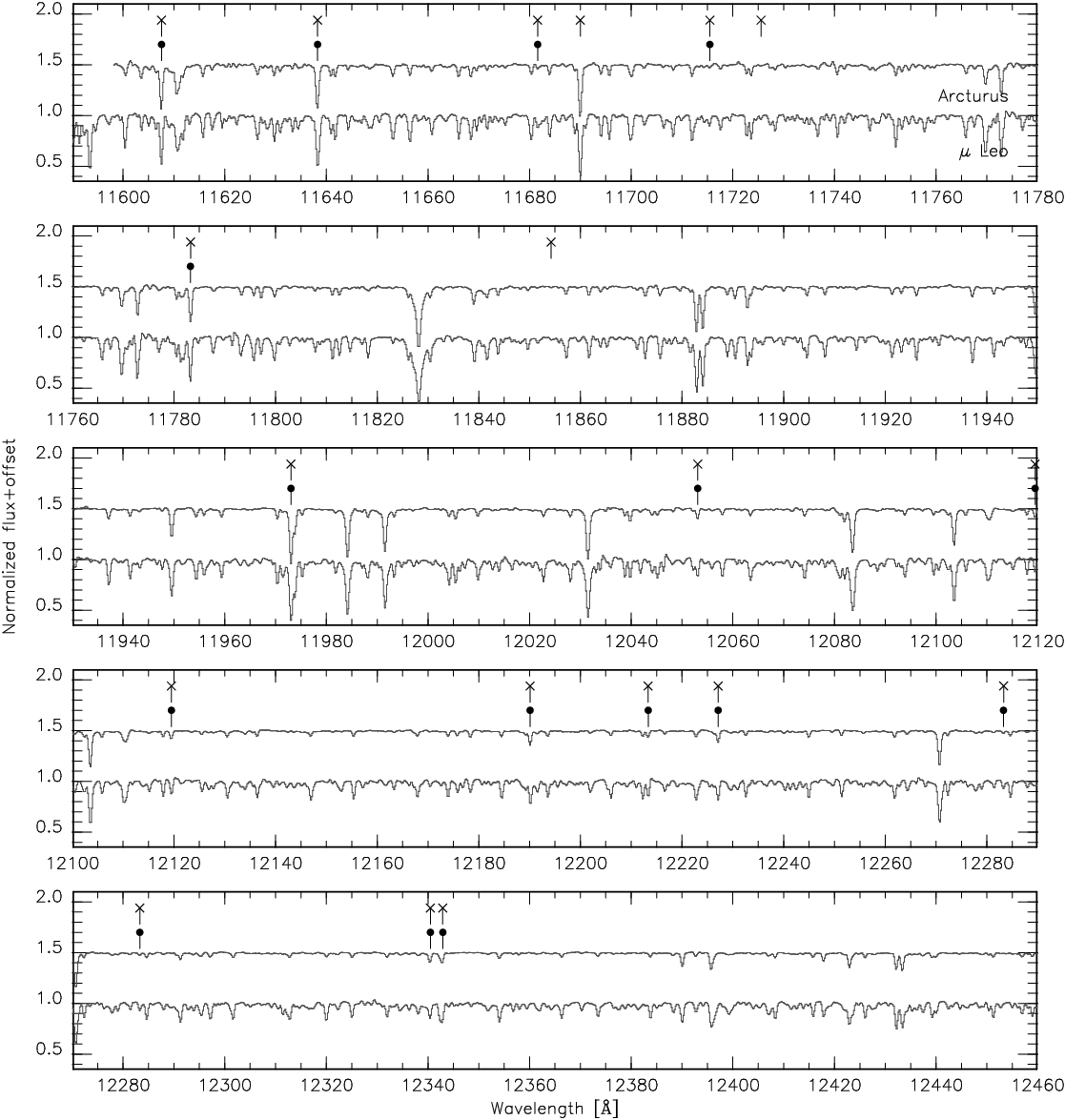}
\end{center}
\caption{continued.}
\end{figure*}
\addtocounter{figure}{-1}
\begin{figure*}
\begin{center}
\includegraphics[bb=80 80 640 640,width=0.98\hsize]{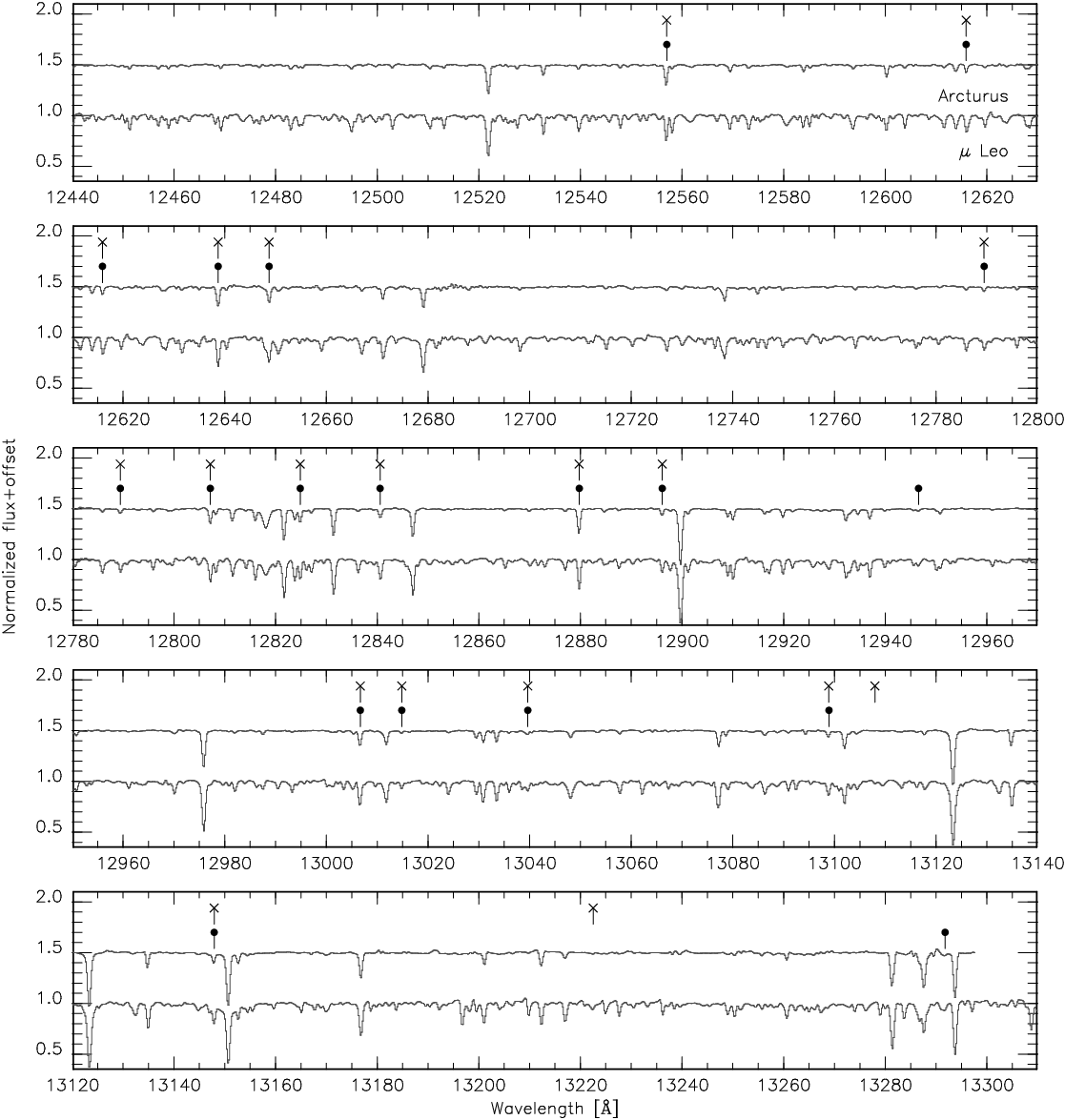}
\end{center}
\caption{continued.}
\end{figure*}

\section{Tools and Basic Data}
\subsection{Atmosphere models and stellar parameters}
\label{sec:atmosphere}

For the abundance analysis, we used SPTOOL
developed by Y.~Takeda (private communication), which utilizes the
ATLAS9/WIDTH9 codes by R.~L.~Kurucz \citep{Kurucz1993}. This tool
synthesizes model spectra using ATLAS9 model atmospheres for a given set
of parameters, including effective temperature ($\Teff$), surface
gravity ($\log g$), and global metallicity ([M/H] or $\logZ$). In these
tools and models, the solar abundance was assumed to be that of
\citet{Anders-1989}. However, in the following discussions, we
translate $\log\epsFe$ values into $\FeH$ by adopting 7.45~dex
\citep{Grevesse-2007} as the solar $\log\epsFe$ value, which was also
adopted in many recent works \citep{Smith-2013,Jofre-2014}.

\begin{figure}
\vspace{3em}
\includegraphics[clip,width=0.98\hsize]{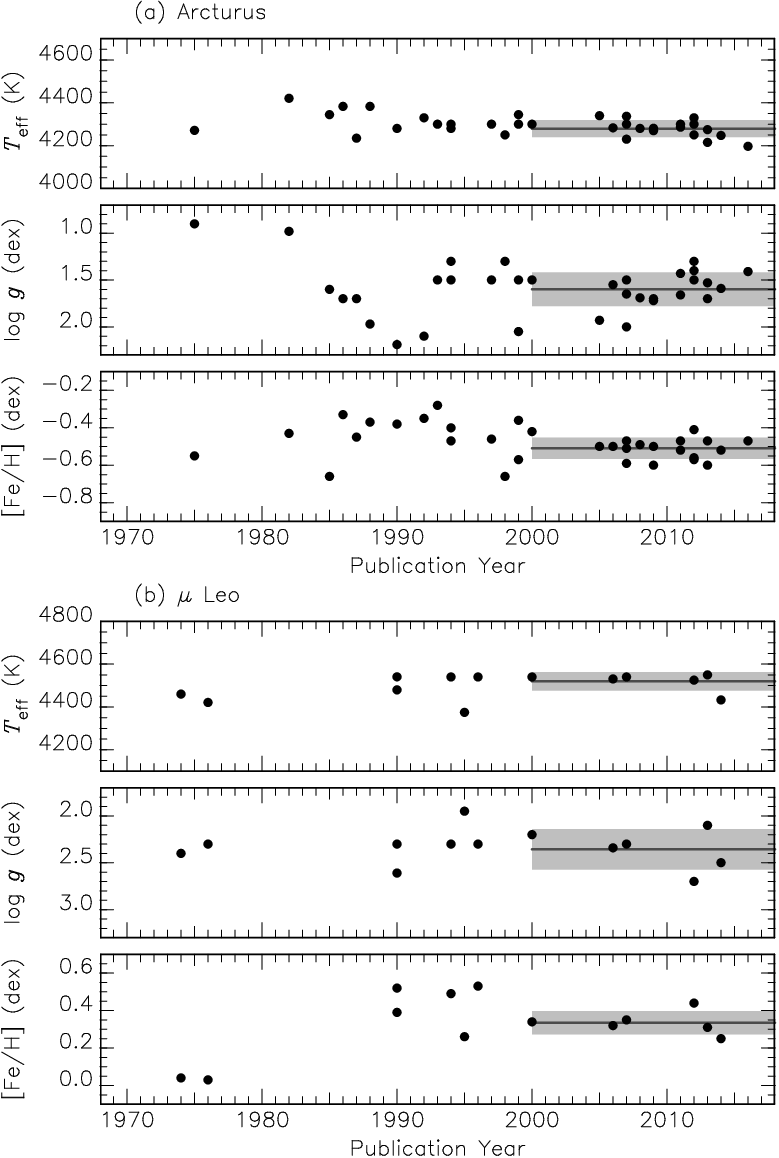}
\caption{
Previous measurements of $\Teff$, $\log g$, and $\FeH$
in the literature.
The upper panels consider 33 papers for Arcturus
\citep{Mackle-1975,Gratton-1982,Bell-1985,Kyrolainen-1986,Leep-1987,Edvardsson-1988,McWilliam-1990,Brown-1992,Peterson-1993,McWilliam-1994,Sneden-1994,Hill-1997,Gonzalez-1998,Tomkin-1999,Thevenin-1999,Carr-2000,Luck-2005,Fulbright-2006,Lecureur-2007,Hekker-2007,Ramirez-2007,Melendez-2008,Worley-2009,Takeda-2009,Ramirez-2011,Bruntt-2011,Sheffield-2012,Britavskiy-2012,Thygesen-2012,Ramirez-2013,Smith-2013,Jofre-2014,Boeche-2016}.
The bottom panels consider 13 papers for $\mu$~Leo
\citep{Oinas-1974,Peterson-1976,McWilliam-1990,Gratton-1990,McWilliam-1994,Luck-1995,Castro-1996,Smith-2000,Fulbright-2006,Lecureur-2007,Thygesen-2012,Smith-2013,Jofre-2014}.
The horizontal line and strip in each panel indicate
the average and standard deviation of the measurements
 made in 2000 or later.
\label{fig2}}
\end{figure}

We adopted the basic stellar parameters and their errors of the two
targets, as listed in Table~\ref{tab1}, from \citet{Heiter-2015}. We
simply use their [Fe/H] values as [M/H] in the atmosphere models. For
comparison, Figure~\ref{fig2} plots previous estimates of $\Teff$, $\log
g$, and $\FeH$, published after 1970, against the publication date. We
included only papers with $\FeH$ in which the assumed solar $\log\epsFe$
was clearly given and those with iron abundance given as $\log\epsFe$,
and all the $\FeH$ values in Figure~\ref{fig2} are scaled with the solar
$\log\epsFe$ of 7.45~dex. The averages of the values published in 2000
or later (18 and 6 papers for Arcturus and $\mu$~Leo, respectively) give
(with standard deviations in parentheses), $\Teff = 4279$~K (40~K),
$\log g=1.60$~dex (0.18~dex), and $\FeH=-0.51$~dex (0.06~dex) for
Arcturus, and $\Teff = 4520$~K (43~K), $\log g=2.36$~dex (0.22~dex), and
$\FeH=+0.33$~dex (0.06~dex) for $\mu$~Leo. These averages agree well
with the parameters from \citet{Heiter-2015} in Table~1.

\subsection{Line lists of VALD3 and MB99}
VALD3 has a large collection of atomic lines, including more than 10,000
\ion{Fe}{1} lines, and molecular lines covering the wavelength range of
the $z^\prime$, $Y$, and $J$ bands. In our spectrum of Arcturus, Ikeda
et~al. (in preparation) identified the atomic lines of various species,
including more than 300 \ion{Fe}{1} lines (see a summary in
\citealt{Taniguchi-2018}). We also considered the line list of MB99,
which includes 363 \ion{Fe}{1} lines in the 1.00--1.34~{$\mu$}m range
among {$\sim$}1000 atomic lines in total. We note that MB99 contains
lines at only $\lambda > 1$~{$\mu$}m and does not cover the entire range
of WINERED spectra. There are 159 lines in both the list of Ikeda
et~al.\ and that of MB99, and there are 475 lines in at least one of the
two lists. The wavelength and the excitation potential (EP in eV) of
each line are consistent between the two line lists. In contrast, the
$\loggf$ values in the two lists are significantly different, as seen
below.

\ion{Fe}{2} lines are not used in our analysis, although there are more
than 10,000 \ion{Fe}{2} lines in VALD3 in the same wavelength range.
MB99 lists 13 \ion{Fe}{2} lines, all of which are also included in
VALD3. We have in fact identified a few \ion{Fe}{2} lines in Arcturus
(to be reported in Ikeda et~al.) and/or $\mu$~Leo. However, most of them
are weaker than 0.01, and none of them is stronger than 0.05 in
depth. Therefore, we focus on abundance measurements using only
\ion{Fe}{1} lines in this paper. We use synthetic spectra for both the
selection of \ion{Fe}{1} lines and the abundance measurements, and we
include all the lines in VALD3 or MB99 (i.e.,\ not only the \ion{Fe}{1}
lines selected in Ikeda et~al.). We use VALD3 for atomic lines when we
consider \ion{Fe}{1} lines and their parameters given in VALD3, and the
same is true for the MB99 lines, in order to avoid mixing the two lists
in our spectral analysis. In both cases, we adopt lines of CN, CO,
C$_2$, CH, and OH molecules in VALD3 because MB99 compiled only atomic
lines.

\section{Selection of \ion{Fe}{1} lines}
\label{sec:selection} To find good \ion{Fe}{1} lines for measuring iron
abundances, we started the line selection from the aforementioned 475
\ion{Fe}{1} lines. First, we excluded 32 lines in the following three
ranges, as they are severely affected by telluric lines:
9,300--9,600~\AA, 11,150--11,600~\AA, and longer than 13,300~\AA. Then,
we measured the depths (minima measured from the normalized continuum)
and central wavelengths of the lines in the synthetic spectra for the
two objects (Arcturus and $\mu$~Leo). We applied the line broadening,
including both macroturbulent and instrumental broadening, corresponding
to $R = 28,000$, for the analysis in this section. If the depth of a
line was shallower than 0.05, the line was rejected. We also rejected
lines that show no minimum in the synthetic spectra for the two objects
within 5~$\kms$ around the expected wavelength. Such lines with a biased
minimum may be strongly blended with other lines. In addition, when two
or more \ion{Fe}{1} lines were detected within 45~$\kms$, we included
only the strongest line if its $X$ value was larger than those of the
other neighboring \ion{Fe}{1} lines by more than 0.5~dex; otherwise, we
rejected both lines. The $X$ index is defined as $X \equiv \loggf -
{\rm EP} \times \theta_{\rm exc}$, where $\theta_{\rm exc} \equiv
5040/(0.86 \times \Teff)$. It is a convenient indicator of line strength
\citep{Magain-1984,Gratton-2006}. In total, 181 (166 in VALD3 and 118 in
MB99) lines in VALD3 and/or MB99 met these criteria.

Then, the impact of blending on each line observed for each object was
examined and used for further selection. We estimated two EWs, $W_1$
and $W_2$, around a target line ($\Lc$) in a synthetic spectrum,
$\fsyn$:
\begin{equation}
W_i = \int_{\Lc-\Delta_i/2}^{\Lc+\Delta_i/2} \{ 1-\fsyn (\lambda)\}
 d\lambda .
\label{eq:EW}
\end{equation}

\begin{figure}
\begin{center}
 
 \includegraphics[width=0.95\hsize]{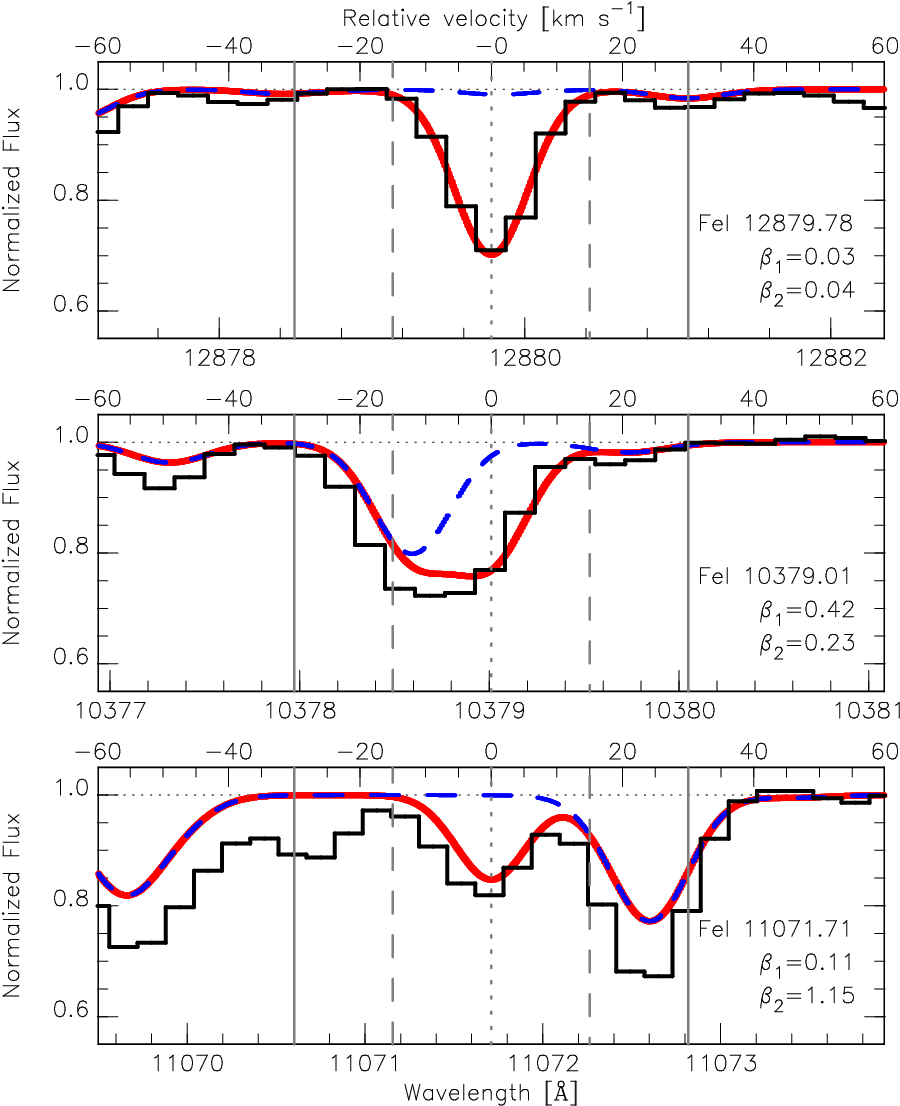}
 \caption{Comparisons between the observational spectrum
 (black) and the two synthetic spectra ($\fsyn$ by red solid curve and
 $\fsynr$ by blue dashed curve; see text for the details) are
 illustrated for three \ion{Fe}{1} lines seen in $\mu$~Leo as examples.
 The MB99 list was used for atomic lines in those synthetic spectra. The
 vertical lines indicate the central wavelength and the velocity ranges corresponding to the widths
 of $\Delta_1=30$ and $\Delta_2=60~\kms$. The three \ion{Fe}{1} lines have
 different $\beta_1$ and $\beta_2$ values from each other, as labeled
 in the panels, and only the one in the top panel was selected for our
 abundance analysis according to the selection criteria of $\beta_1<0.3$
 and $\beta_2<1$.\label{fig_profile}}
\end{center}
 \end{figure}

For the EW of the target line itself and contaminations of lines in
neighboring wavelengths, we consider two different integration ranges,
$\Delta_1$ and $\Delta_2$, which correspond to velocities of 30 and
60~$\kms$, respectively. Neighboring lines other than the target line
can also contribute to these EWs ($W_1$ and $W_2$). In addition, to
evaluate the contamination, we constructed synthetic spectra, $\fsynr$,
with the target \ion{Fe}{1} line removed from the line lists for each of
the two stars. The EW of contaminating lines, $W_i^\dagger$, can be
estimated by considering Equation~(\ref{eq:EW}) but with $\fsyn$
replaced by $\fsynr$. Combining these EWs, we consider two indices,
\begin{eqnarray}
\beta_1 &=& W_1^\dagger / W_1 , \\
\beta_2 &=& (W_2^\dagger-W_1^\dagger) / W_1 ,
\end{eqnarray}
as indicators of blending. The former measures the contamination to the
main part of each target line, and the latter measures the contamination
mainly to the continuum part around the line. Firstly, we rejected lines
for which $\fsynr - \fsyn$ does not exceed 0.05. The 181 lines were
selected because they are deeper than 0.05 in $\fsyn$ in the previous
stage, but we found that a significant number of them are deep because
of the contamination. Among the 118 lines in MB99, for example, 53 and
25 were rejected in the cases of Arcturus and $\mu$~Leo, respectively,
considering the depths in $\fsynr - \fsyn$. Then, we rejected lines
with $\beta_1 > 0.3$ or $\beta_2 > 1$; 8 and 21 lines were rejected in
the cases of Arcturus and $\mu$~Leo, respectively, although those lines
are strong enough. Figure~\ref{fig_profile} shows examples of
\ion{Fe}{1} lines with different $\beta_1$ and $\beta_2$ values. We
note that the selection in this section was made on the basis of
synthetic spectra, and not observed ones. Some \ion{Fe}{1} lines look
isolated enough in synthetic spectra but turn out to be severely blended
with neighboring strong lines that are not reproduced in the synthetic
spectra (see Section \ref{sec:bootstrap}). All of the above mentioned
rejections were made independently for each combination of the line list
(VALD3 or MB99) and the object (Arcturus or $\mu$~Leo).

Tables~\ref{tab2} and \ref{tab3} list the selected lines, and
Table~\ref{tab4} lists the number of the lines, $N_1$, for each
combination of line list and object. Some lines were selected only for
one of the two objects owing to the large difference in
metallicity. Among the 97 selected lines from VALD3 (Table~\ref{tab2}),
24 lines are weak only in Arcturus, while there are no lines, as
expected, which are weak in $\mu$~Leo but strong enough in Arcturus. In
contrast, 6 lines were rejected due to the blending for $\mu$~Leo only,
and no lines selected for $\mu$~Leo show strong blends in Arcturus. The
situation is similar for the 75 selected lines from MB99. 18 lines were
rejected for Arcturus because they are weak in $\fsynr-\fsyn$, while no
line selected for Arcturus is weak in $\mu$~Leo. 3 lines were rejected
because of blends in $\mu$~Leo, but no line selected for $\mu$~Leo was
rejected owing to blends in Arcturus.

\begin{deluxetable}{ccccc}
\tabletypesize{\small}
\tablecaption{List of \ion{Fe}{1} lines selected from VALD3 and abundances\label{tab2}}
\tablehead{
\colhead{Wavelength}
& \colhead{EP}
& \colhead{$\log gf$}
& \colhead{Arcturus}
& \colhead{$\mu$~Leo}
\\ 
\colhead{(\AA)}
& \colhead{(eV)}
& \colhead{(dex)}
& \colhead{(dex)}
& \colhead{(dex)}
}
\startdata
9117.1309 & 2.8581 & $-3.454$ & 6.970 & 7.888\\ 
9118.8806 & 2.8316 & $-2.115$ & 6.411 & 8.612\\ 
9146.1275 & 2.5881 & $-2.804$ & 6.828 & 6.749\\ 
9210.0240 & 2.8450 & $-2.404$ & 6.789 & 7.276\\ 
9602.1301 & 5.0117 & $-1.744$ & (w) & 7.408\\ 
9653.1147 & 4.7331 & $-0.684$ & 6.780 & 7.545\\ 
9657.2326 & 5.0856 & $-0.780$ & 6.768 & 7.152\\ 
9738.5725 & 4.9913 & $+0.150$ & 6.861 & 7.308\\ 
9753.0906 & 4.7955 & $-0.782$ & 6.850 & (*) \\ 
9791.6983 & 2.9904 & $-4.223$ & (w) & 7.126\\ 
9800.3075 & 5.0856 & $-0.453$ & 6.558 & 7.457\\ 
9811.5041 & 5.0117 & $-1.362$ & 7.100 & 7.646\\ 
9820.2408 & 2.4242 & $-5.073$ & (w) & (*) \\ 
9861.7337 & 5.0638 & $-0.142$ & 6.647 & (b) \\ 
9868.1857 & 5.0856 & $-0.979$ & 7.098 & 8.246\\ 
9889.0351 & 5.0331 & $-0.446$ & 6.974 & 7.660\\ 
9937.0898 & 4.5931 & $-2.442$ & (w) & 7.544\\ 
9944.2065 & 5.0117 & $-1.338$ & 7.046 & 7.401\\ 
9980.4629 & 5.0331 & $-1.379$ & 6.851 & 7.935\\ 
10041.472 & 5.0117 & $-1.772$ & (w) & 7.958\\ 
10065.045 & 4.8349 & $-0.289$ & 6.774 & 7.618\\ 
10081.393 & 2.4242 & $-4.537$ & 6.995 & 7.602\\ 
10114.014 & 2.7586 & $-3.692$ & 6.918 & (b) \\ 
10145.561 & 4.7955 & $-0.177$ & 6.947 & (b) \\ 
10155.162 & 2.1759 & $-4.226$ & 6.770 & 7.459 
\enddata
\tablenotetext{}{These are the first 25 lines. 
 Lines weaker than the limit 0.05 in depth in synthetic
 spectra, are flagged as
(w), and lines that are blended too much are flagged as (b). The flag
 (*) indicates lines whose
 abundance could not be obtained or was rejected. See the details of the
 line selection in text. The entire list is
 available as an ASCII file in the online journal.}
\end{deluxetable}

\begin{deluxetable}{ccccc}
\tabletypesize{\small}
\tablecaption{List of \ion{Fe}{1} lines selected from MB99 and abundances\label{tab3}}
\tablehead{
\colhead{Wavelength}
& \colhead{EP}
& \colhead{$\log gf$}
& \colhead{Arcturus}
& \colhead{$\mu$~Leo}
\\ 
\colhead{(\AA)}
& \colhead{(eV)}
& \colhead{(dex)}
& \colhead{(dex)}
& \colhead{(dex)}
}
\startdata
10019.79 & 5.48 & $-1.44$ & (w) & 7.582\\ 
10032.86 & 5.51 & $-1.36$ & (w) & 7.522\\ 
10041.47 & 5.01 & $-1.84$ & (w) & 7.982\\ 
10065.05 & 4.84 & $-0.57$ & 7.144 & 7.825\\ 
10081.39 & 2.42 & $-4.53$ & 6.963 & 7.459\\ 
10114.02 & 2.76 & $-3.76$ & 7.010 & (b) \\ 
10145.57 & 4.80 & $-0.41$ & 7.335 & 8.342\\ 
10155.16 & 2.18 & $-4.36$ & 6.901 & 7.438\\ 
10167.47 & 2.20 & $-4.26$ & 7.071 & 7.757\\ 
10195.11 & 2.73 & $-3.63$ & 6.915 & 7.800\\ 
10216.32 & 4.73 & $-0.29$ & 7.262 & 8.006\\ 
10218.41 & 3.07 & $-2.93$ & 7.092 & 8.038\\ 
10230.78 & 5.87 & $-0.70$ & (w) & 7.774\\ 
10265.22 & 2.22 & $-4.67$ & 6.962 & 7.416\\ 
10307.45 & 4.59 & $-2.45$ & (w) & 7.524\\ 
10340.89 & 2.20 & $-3.65$ & 7.092 & 7.508\\ 
10347.96 & 5.39 & $-0.82$ & 6.970 & 8.024\\ 
10353.81 & 5.39 & $-1.09$ & (w) & 7.707\\ 
10395.80 & 2.18 & $-3.42$ & 6.749 & 7.353\\ 
10401.72 & 3.02 & $-4.36$ & (w) & 7.583\\ 
10435.36 & 4.73 & $-2.11$ & (w) & 7.852\\ 
10452.75 & 3.88 & $-2.30$ & 6.781 & 7.713\\ 
10469.66 & 3.88 & $-1.37$ & 6.984 & 7.908\\ 
10532.24 & 3.93 & $-1.76$ & 7.151 & 7.733\\ 
10555.65 & 5.45 & $-1.39$ & (w) & 7.565 
\enddata
\tablenotetext{}{These are the first 25 lines. The entire list is
 available as an ASCII file in the online journal. The
 meanings of
 the flags, (w), (b), and (*), are same as in Table~\ref{tab2}.}
\end{deluxetable}

\begin{deluxetable}{cccccc}
\tabletypesize{\small}
\tablecaption{Microturbulence and iron abundance\label{tab4}}
\tablehead{
\colhead{Line list}
& \colhead{$N_1$}
& \colhead{$N_2$}
& \colhead{$\xi$}
& \colhead{$\log\epsFe$}
& \colhead{$r$}
\\ 
\colhead{}
& \colhead{}
& \colhead{}
& \colhead{($\kms$)}
& \colhead{(dex)}
& \colhead{}
}
\startdata
\hline
\multicolumn{6}{c}{Arcturus} \\
 VALD3 & 73 & 67 & $1.22^{-0.12}_{+0.12}$ & $6.81^{+0.06}_{-0.06}$ & $-0.946$ \\ 
 MB99 & 57 & 53 & $1.20^{-0.11}_{+0.11}$ & $7.01^{+0.04}_{-0.04}$ & $-0.875$ \\ 
\hline
\multicolumn{6}{c}{$\mu$~Leo} \\
 VALD3 & 91 & 79 & $1.16^{-0.24}_{+0.23}$ & $7.62^{+0.11}_{-0.10}$ & $-0.909$ \\ 
 MB99 & 72 & 63 & $1.54^{-0.17}_{+0.17}$ & $7.73^{+0.06}_{-0.05}$ & $-0.828$ 
\enddata
\end{deluxetable}

\section{Measurement of microturbulence and metallicity}
\subsection{Bootstrap method to measure $\xi$ and $\log\epsFe$}
\label{sec:bootstrap}

We determined the iron abundance ($\log\epsFe$) and the microturbulence
($\xi$) simultaneously for each combination of object (Arcturus or
$\mu$~Leo) and line list (VALD3 or MB99) as follows. The basic
assumption of the method is that the $\log\epsFe$ values should be
independent of line strength, as is often assumed in the classical
method of abundance analysis (see the Introduction).

We measured the $\log\epsFe$ of each \ion{Fe}{1} line for 21 different
$\xi$ values from 0.5 to 2.5~$\kms$ with a step of 0.1~$\kms$. For each
combination of line and $\xi$, $\log\epsFe$ was estimated by a
least-squares fit to a small part of the spectrum around the line using
MPFIT \citep{Takeda-1995}, which is implemented in SPTOOL. Each MPFIT
run was performed with a fixed $\xi$. We used a fitting window,
$[\lambda_c-\Delta_2/2:\lambda_c+\Delta_2/2]$, where $\lambda_c$ is the
central wavelength of each line and $\Delta_2$ is the wavelength shift
corresponding to a redshift of $60~\kms$, as
Equation~(\ref{eq:EW}). MPFIT searches for an optimized solution by
treating the following as free parameters: $\log\epsFe$, the width of
Gaussian line broadening (including macroturbulence and instrumental
broadening), and a small wavelength offset $\Delta\lambda$, which
compensates for any remaining errors in the wavelength calibration and
in the correction of the redshift of the target. We thus obtained
$\log\epsFe$ values for the grid of 21 $\xi$ values for individual
\ion{Fe}{1} lines. The number of lines measured for each combination of
line list and object is given as $N_2$ in Table~\ref{tab4}. Note that
MPFIT failed to give a solution for a few lines for $\mu$~Leo, namely,
\ion{Fe}{1} 11026.78, 11053.52, and 11135.96~{\AA} from both line lists,
\ion{Fe}{1} 9753.09, 9820.24, 13145.07, and 11119.80~{\AA} from VALD3
only, and \ion{Fe}{1} 11715.49, and 13291.78 ~{\AA} from MB99
only. Visual inspection of its observed spectrum around these lines
suggests that they are blended by one or two other strong lines. Such
cases could have been rejected based on the $\beta_1$ and $\beta_2$
indices, but the blends around the above lines were not reproduced by
the synthetic spectra (on the basis of MB99 and VALD3). Four and three
of these lines were rejected for Arcturus when we used VALD3 and MB99,
respectively, because they were predicted to be weak, but for the other
lines we obtained $\log\epsFe$ of Arcturus. In Tables~\ref{tab2} and
\ref{tab3}, we include these lines for which MPFIT failed, marked with
an asterisk (*), because they may still be useful in some cases or once
the line lists have been improved to reproduce the spectra including the
neighboring lines. The \ion{Fe}{1} line at 13291.78~{\AA} in the MB99
list was selected for $\mu$~Leo; however MPFIT gives completely wrong
$\log\epsFe$ values, higher than 10~dex. We found that this line is
severely blended in the observed spectrum of $\mu$~Leo, but it looks
fairly isolated in the synthetic spectrum. This inconsistency probably
causes the absurd $\log\epsFe$ values. We, therefore, reject the MPFIT
measurements of this line but include the line in Table \ref{tab3}
marked with the asterisk (*). These rejected lines are not included in
$N_2$ in Table~\ref{tab4}. Additionally, the lines with $X>-6$ are not
used when we estimate the final iron abundances
(Section~\ref{sec:twolists}), and those lines are not included in $N_2$
in the table.

We then used a bootstrap method to obtain not only the best estimates of
$\xi$ and $\log\epsFe$ but also respective errors. We repeatedly
extracted $N_2$ randomly-selected lines among the $N_2$ lines with
$(\xi, \log\epsFe)$ available. Note that for each bootstrap sample, each
line may be selected more than once and some lines may be excluded.

\begin{figure}
 \begin{center}
 \includegraphics[width=0.98\hsize]{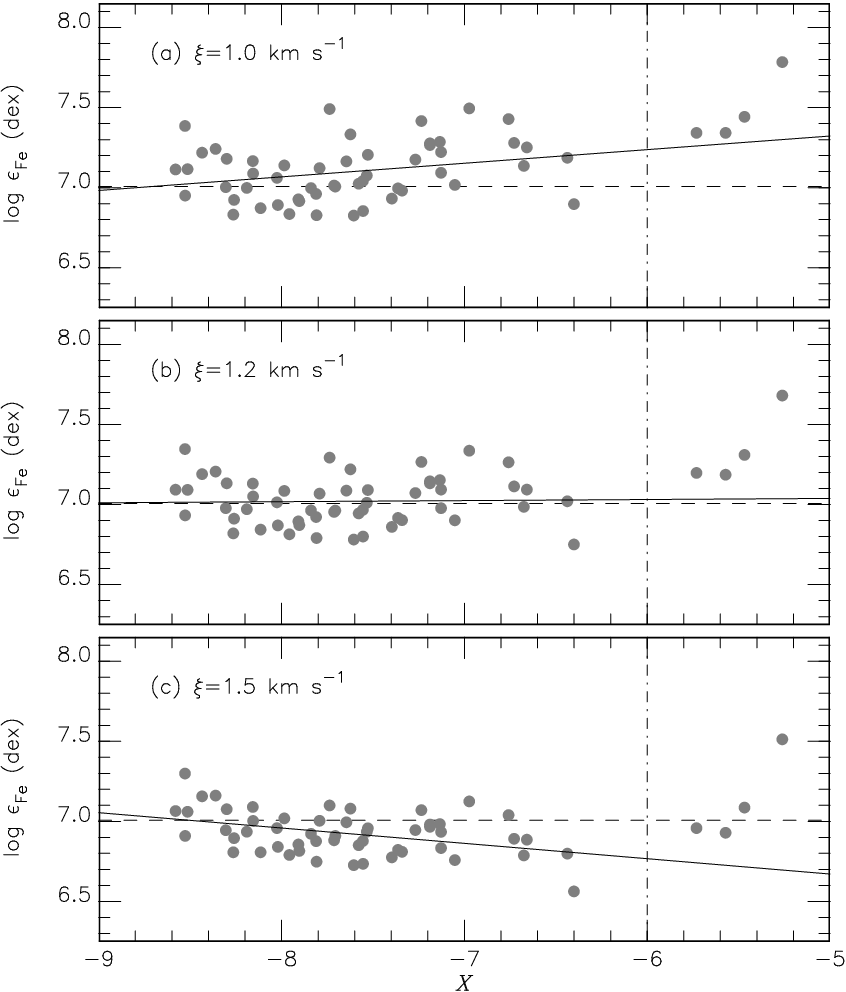}
\caption{Dependency of $\log\epsFe$ on line strength indicated by $X$
at different $\xi$ values, (a)~1.0, (b)~1.2,
 and (c)~1.5~$\kms$. The
 solid line in each panel shows the linear fit to the $(X, \log\epsFe)$ points at $X<-6$.
This plot is for the combination of the MB99 line list and
 Arcturus. The \ion{Fe}{1} lines with $X>-6$,
 indicated by the vertical line in each panel, were not used in the final $(\xi, \log\epsFe)$ determination. \label{fig3}}
 \end{center}
 \end{figure}

For a given set of the $(\xi, \log\epsFe)$ values for a bootstrap
sample, we obtained the best estimates of $\xi$ and $\log\epsFe$ as
follows. First, we searched for $\xi$ that leads to no trend of
$\log\epsFe$ of individual lines against the line strength. We
considered the $X$ value introduced in Section~\ref{sec:selection} as a
proxy of the line strength, and made a simple least-squares fit,
\begin{eqnarray}
\log\epsFe = aX + b, \label{eq:X_logeps}
\end{eqnarray}
to calculate the trend, $a$, for each $\xi$ of the grid.
Figure~\ref{fig3} illustrates that lines with different strengths have
different responses to $\xi$. Lines with large $X$ values, but within
the range of $X<-6$, tend to give smaller $\log\epsFe$ for larger $\xi$.
This leads to a monotonic decrease in the slope $a$ with increasing
$\xi$. One can, thus, find a $\xi$ that gives $a=0$ by interpolating
two neighboring $\xi$ values where $a$ turns from positive to negative.
In Figure~\ref{fig3}, $a$ is almost zero at $\xi=1.2~\kms$ (panel
b). The lines at $X>-6$ are biased toward higher $\log\epsFe$ values,
and we will discuss their impact on the estimate of $\xi$ and
$\log\epsFe$ in Section~\ref{sec:sigmas}. For the $\xi$ obtained, we
calculated $\log\epsFe$ for $N_2$ individual lines of the bootstrap
sample by interpolating the grid points of $(\xi, \log\epsFe)$ and took
the average of the $\log\epsFe$ values. This gives the best estimate of
$(\xi, \log\epsFe)$ for the given bootstrap sample. We then took the
median and also the 16th and 84th percentiles (as the $\pm 1\,\sigma$
range) in each of the histograms of $\xi$ and $\log\epsFe$ values
obtained after a large number of bootstrap samples. We repeated this
procedure one million times ($N_{\rm b}=1,000,000$) in this study, and
the best estimates of $(\xi, \log\epsFe)$ are listed in Table~\ref{tab4}
for each combination of the line list and object. We also calculated
the correlation coefficient of the two parameters,
\begin{eqnarray}
\label{eq:r}
r = \frac{\sum \left(\xi^i - \langle \xi \rangle \right) \left( \log\epsFe ^i - \langle \log\epsFe \rangle \right)}{\sqrt{\sum \left(\xi^i - \langle \xi \rangle \right)^2} \sqrt{\sum \left( \log\epsFe ^i - \langle \log\epsFe \rangle \right)^2}}
\end{eqnarray}
where $\xi^i$ and $\log\epsFe^i$ are the microturbulence and iron abundance obtained
for each bootstrap sample, and $\langle \xi \rangle$ and $\langle \log\epsFe \rangle$
are their means (not medians). Each of the summations in Equation~(\ref{eq:r}) takes the integer $i$ for $N_b$ lines, i.e.,~$1\leq i\leq N_{\rm b}$.

\begin{figure}
 \begin{center}
 \includegraphics[width=0.98\hsize]{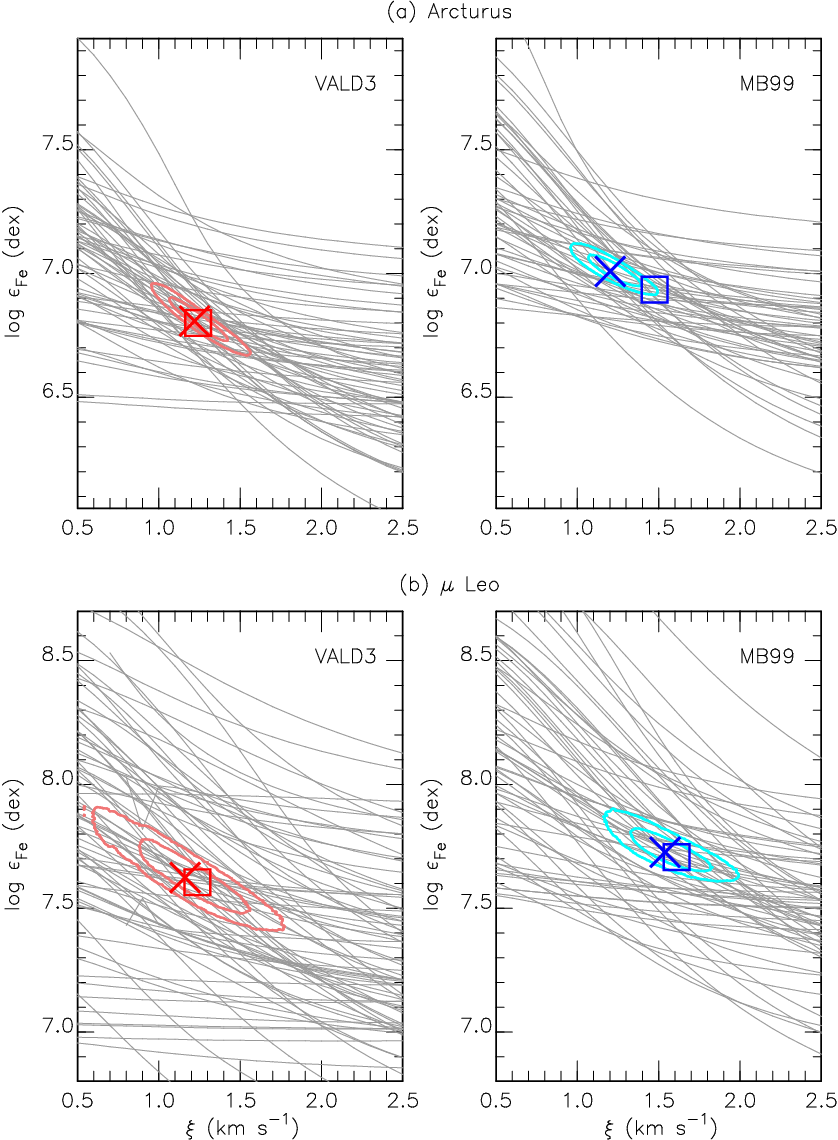}
\caption{Contours for the density distribution of $(\xi, \log\epsFe)$
obtained in the bootstrap simulation. The inner and outer contours show
the ranges that include 68.26\,\% ($1\,\sigma$) and 95.44\,\% ($2\,\sigma$)
of the 1,000,000 bootstrap samples. Four panels are given for
the combinations of line lists (VALD3 and MB99) and
 targets (Arcturus and $\mu$~Leo). In each panel,
gray curves indicate the dependency of $\log\epsFe$ for individual
 lines. The cross symbol indicates the best estimates that we obtained
for each set (Table~\ref{tab4}), and the open square indicates
the estimates obtained with strong lines with $X>-6$ included
(see Section~\ref{sec:twolists}).
 \label{fig4}}
 \end{center}
 \end{figure}

The contours in Figure~\ref{fig4} represent the distribution
of $(\xi, \log\epsFe)$ obtained in the bootstrap simulation.
The large $N_{\rm b}$ was used mainly to obtain smooth contours in Figure~\ref{fig4},
although we could obtain reasonably stable values
including $1\,\sigma$ confidence intervals at around $N_{\rm b}=10,000$.
 There is a linear anticorrelation, as expected, between $\xi$ and $\log\epsFe$,
which shows that the errors in the two parameters are anticorrelated (see $r$ in Table~\ref{tab4}). 
We do not use $r$ later in this paper, but it is a useful
indicator of how much the measured $\log\epsFe$ depends on
the $\xi$ estimated.
For example, $r$ is expected to vary with the proportion of strong lines.
Using more weak lines would reduce the anticorrelation because the $\log\epsFe$ values of weak lines 
have a smaller dependency on $\xi$.

 \begin{figure}[!htb]
 \begin{center}
 
 \includegraphics[clip,width=0.98\hsize]{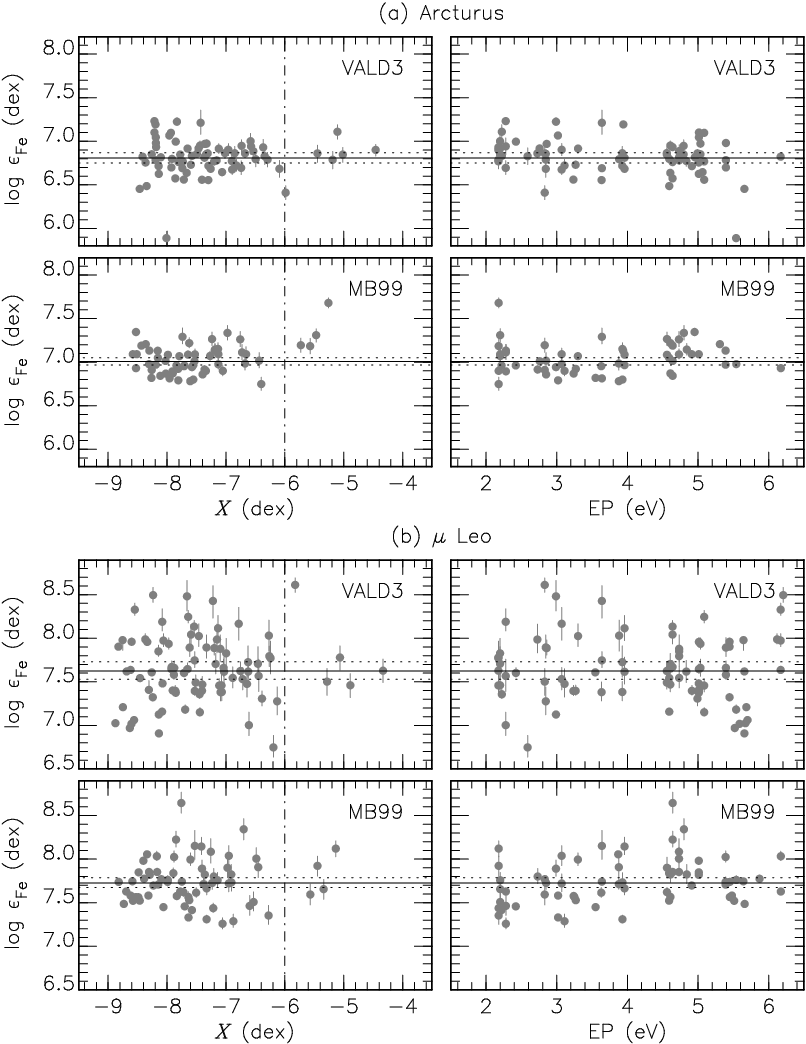}
\caption{The $\log\epsFe$ values obtained for individual \ion{Fe}{1} lines
are plotted against the line strength indicator, $X$,
on the left-hand side and against the excitation potential, EP,
on the right-hand side. For each of the two targets
(Arcturus in the upper panels and $\mu$~Leo in the lower panels),
the results for the two line lists (VALD3 and MB99) are presented.
The horizontal solid line and the dashed lines in each panel
indicate the best estimate and $1\,\sigma$ confidence intervals
for the combination of line list and target. The
 \ion{Fe}{1} lines with $X>-6$, indicated by the vertical line in each panel, were not used in the final $(\xi, \log\epsFe)$ determination.
\label{fig5}}
 \end{center}
 \end{figure}

Now, we estimate $\log \epsFe$ values of individual lines with the best
estimates of $\xi$ that are given in Table~\ref{tab4}. For each
combination of object and line list, each \ion{Fe}{1} line has 21
measurements of $\log\epsFe$ at different $\xi$ values, and we
interpolated $\log\epsFe$ values at the two grid points of $\xi$ next to
its best estimate. The $\log\epsFe$ values obtained for individual lines
are listed in Table~\ref{tab2} for VALD3 and in Table~\ref{tab3} for
MB99. In the two tables, lines weaker than the limit are flagged as (w),
and lines that are blended too much are flagged as (b). Lines whose
MPFIT measurements were unavailable or rejected were not used for the
abundance analysis, but we include them in the tables with the (*)
flag. Figure~\ref{fig5} plots the individual $\log\epsFe$ values against
the $X$ value and EP. For both objects, the $X$ values of the measured
lines are spread over a wide range, approximately between $-9$ and
$-5$~dex. Such a wide range among the lines in the $z^\prime$, $Y$, and
$J$ bands is advantageous, for example, compared with a narrow range,
$-8.3$ to $-7.3$~dex, covered by the $H$-band lines used by
\citet{Smith-2013}. The $\log\epsFe$ shows little dependency on $X$ as
demanded in the analysis and also have no clear dependency on EP,
indicating that the adopted $\Teff$ are reasonable. The scatters of
$\log\epsFe$ from individual lines are larger for $\mu$~Leo than for
Arcturus. This is probably because the spectrum of $\mu$~Leo has
stronger contaminating lines, especially CN lines, than Arcturus
\citep{McWilliam-1994,Smith-2013}, which makes it harder to trace the
continuum.

\subsection{Comparison between the two line lists}
\label{sec:twolists}
There are a few differences between the estimates
of $(\xi, \log\epsFe$) obtained with the two line lists. 

Firstly, in Table~\ref{tab4}, the standard errors for $\log\epsFe$ from the two lists
are similar to each other for Arcturus. The number of \ion{Fe}{1} lines is
larger for VALD3, but the measured $\log\epsFe$ has a slightly larger scatter
than for MB99, which is compensated by the larger $N_2$.
For $\mu$~Leo, the scatter of $\log\epsFe$ is rather large
with VALD3 (Figure~\ref{fig5}), and this leads to a larger standard error even with
a larger number of \ion{Fe}{1} lines.

\begin{figure}[!tbh]
 \begin{center}
 \includegraphics[clip,width=0.98\hsize]{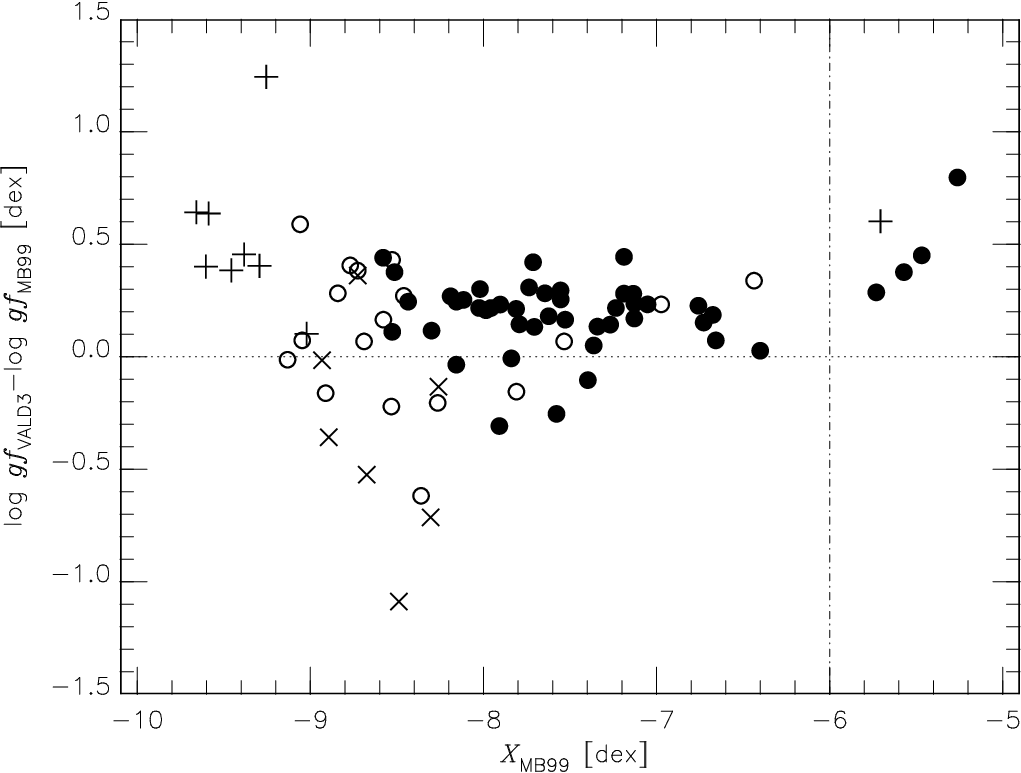}
 \caption{Comparison of
the $\loggf$ values for the two line lists VALD3 and MB99. The filled
circles indicate the lines used for all combinations of line list and
target, and the open circles indicate those used for both line lists but
for only one of the two targets, Arcturus or
 $\mu$~Leo. The $+$ and 
$\times$ symbols indicate the lines used in only one of
the line lists ( $+$ for VALD3 and
 $\times$ for MB99) for both targets. The
temperature of Arcturus, $\Teff =4286$~K, and the $\log gf$ values for
MB99 are used for calculating the $X$ values (abscissa).\label{fig7}}
 \end{center}
 \end{figure}

Secondly, the resultant $\log\epsFe$ values for MB99 are slightly higher
than those obtained for VALD3. In fact, there is a systematic offset in
the $\loggf$ values between the two line lists (Figure~\ref{fig7}). The
systematic offset, $\sim$0.2~dex, approximately corresponds to the
difference in $\log\epsFe$ for Arcturus obtained with VALD3 and MB99. In
contrast, the corresponding difference in the case of $\mu$~Leo is smaller.
 Although the offsets
in the $\loggf$ have a direct impact on the $\log\epsFe$ estimation, the
different $\xi$ values obtained for $\mu$~Leo with the two lists (larger
$\xi$ with MB99 than VALD3) partly compensate for
this systematic offset.

Finally, the final estimates depend slightly on whether
very strong lines with $X>-6$ are used or not. In Figure~\ref{fig7},
very strong lines clearly show a systematic tilt.
These strong lines have an impact on the slopes, e.g., seen in Figure~\ref{fig3}.
The lower $\loggf$ values of the stronger lines in MB99
would give higher $\log\epsFe$ values with a fixed $\xi$,
but this would also cause a tilt in Figure~\ref{fig3}.
A larger $\xi$ is therefore required so that $\log\epsFe$ values of
strong and weak lines get balanced.
 While this is an important difference between
the two line lists, generally speaking, it is suggested that
using very strong lines often introduces complications such as
non-LTE effects into a chemical abundance analysis
\citep[e.g.,][]{Kovtyukh-1999,Gratton-2006,Takeda-2013}. 
Based on synthetic spectra, we found that, in case of lines with X $\gtrsim-6
$, the line core does not grow any more
with increasing metallicity and the damping wing starts to contribute to the EW
at around the solar metallicity. 
If we run the bootstrap method with the same lines but 
including those with $X>-6$, we obtain moderately different results
for the MB99 list, as illustrated in Figure~\ref{fig4}.
 Four lines from MB99 
have $X>-6$, and including them leads to
 higher $\xi$ and 
lower $\log\epsFe$ values:
$(\xi, \log\epsFe)=$$(1.47{\pm 0.18}, 6.94{\mp 0.05})$ for Arcturus
and $(1.61{\pm 0.16},
7.71{\mp 0.06})$ for $\mu$~Leo.
The changes caused by including the strongest lines are marginally significant, 1--2\,$\sigma$, for the former
but are negligible for the latter.
Figure~\ref{fig5} shows that one line, \ion{Fe}{1} $\lambda$~11973.04,
 with the
largest $(X,\log \epsFe)$ has
a particularly strong impact on the slope in the $X$ versus $\log\epsFe$
diagram for Arcturus with MB99.
The same line gives $\log\epsFe$ $\sim$8.10~dex, which is also higher than
the average, for $\mu$~Leo. However, the scatter of $\log\epsFe$ from lines
within the low-$X$ range is large, which explains the relatively small
effect of including the high-$X$ lines for $\mu$~Leo.
In contrast, six VALD3 lines that we
 selected have $X>-6$, but 
including them has a negligible impact on
the $(\xi, \log\epsFe)$ measurements. 
For VALD3, the \ion{Fe}{1} $\lambda$~11973.046 line leads to $\log\epsFe$ values that are very close to
the average abundances from other lines for both Arcturus and
$\mu$~Leo. This line corresponds to the rightmost point in
Figure~\ref{fig7} and has a very large difference, 0.8~dex, between the $\log gf$ values
in the two line lists. Considering these complications, we decided to
adopt the $(\xi, \log\epsFe)$ values obtained without the lines at
$X>-6$ as our best estimates.
Although the $\log \epsFe$ from individual lines depend on $\xi$ as described above, we found that
the [Fe/H] obtained in different works are
not correlated with $\xi$ (Figure 8). This is probably
because systematic differences in previous works,
such as differences in line lists and atmosphere models, introduced a
scatter larger than the expected correlation between the two parameters.

\subsection{Effects of stellar parameters on metallicity}
\label{sec:sigmas}

Here, we estimate how much the uncertainties in the stellar parameters, $\Teff$, $\log g$, and [M/H], affect the estimates of $\log\epsFe$.
We adopt the errors in these parameters from
\citet{Heiter-2015}, as given in Table~\ref{tab1}.
To evaluate the effect of changing the three parameters,
we added positive and negative offsets
to each parameter in the atmosphere models one by one.
For each offset, we ran MPFIT and obtained $\log\epsFe$
for the $N_2$ lines and calculated their means. We did not use
the bootstrap method described in Section \ref{sec:bootstrap} for this step
because we need to estimate the effect of a parameter at a fixed $\xi$.
Then, we compared the above means with the counterparts
of the mean $\log\epsFe$ with the stellar parameters in Table~\ref{tab1}. 
This gives the offsets in $\log \epsFe$, $\Delta (\Teff)$,
$\Delta (\log g)$, and $\Delta ({\rm [M/H])}$,
as a result of changing the stellar parameters (Table~\ref{tab5}).

For both objects and for both line lists, we found that varying the
temperature or the gravity gives rather tiny changes in $\log\epsFe$.
Synthetic spectra with the same parameters but an offset of 50~K in
$\Teff$ or an offset of $\pm$~0.1~dex in $\log g$ do not actually
 show any noticeable changes in the
\ion{Fe}{1} lines. The $\Delta ({\rm
[M/H]})$ is larger compared with these two. 
The $\Delta ({\rm [M/H]})$ of Arcturus is smaller than
that of $\mu$~Leo. We believe that this is simply because 
the $\sigma {\rm [M/H]}$ of Arcturus is smaller than
that of $\mu$~Leo.
We combine the
$\Delta $ values with the confidence intervals of $\log\epsFe$
estimated by the bootstrap method, the $\Delta_{\rm b}$, in Table~\ref{tab5}. Note
that the $\Delta_{\rm b}$
correlated with $\xi$ include other errors,
e.g., observational errors in the
spectra and errors in $\loggf$. Combining the above errors, we can
estimate the total error as
\begin{eqnarray} 
 \Delta_{\rm total}
 = \sqrt{\Delta _{\rm b}^{2}+{\Delta (\Teff)}^{2}+{\Delta (\log
 g)}^{2}+{\Delta ({\rm [M/H]})}^{2}},
\end{eqnarray}
which is given in Table~\ref{tab5}. 
Here, we ignored the covariant terms.
The previous estimates that we compiled in Figure~\ref{fig2} show
no clear correlation between any two of the four parameters,
$\Teff$, $\log g$, ${\rm [Fe/H]}$, or $\xi$.

\begin{deluxetable*}{crrrrrrrrrr}[!htb]
 \tabletypesize{\small}
\tablecaption{Effects of stellar parameters on iron abundance\label{tab5}}
\tablehead{
\colhead{Line list}
& \colhead{$\sigma \Teff$}
& \colhead{$\Delta ({\Teff})$}
& \colhead{$\sigma \log g$}
& \colhead{$\Delta ({\log g})$}
& \colhead{$\sigma {\rm [M/H]}$}
& \colhead{$\Delta ({\rm [M/H]})$}
& \colhead{$\sigma_\xi$}
& \colhead{${\Delta_{\rm b}}^{+}$}
& \colhead{${\Delta_{\rm b}}^{-}$}
& \colhead{${\Delta_{\rm total}}$}
\\ 
\colhead{}
& \colhead{(K)}
& \colhead{(dex)}
& \colhead{(dex)}
& \colhead{(dex)}
& \colhead{(dex)}
& \colhead{(dex)}
& \colhead{$(\kms)$}
& \colhead{(dex)}
& \colhead{(dex)}
& \colhead{(dex)}
}
 \startdata
\multicolumn{10}{c}{Arcturus} \\
VALD3 & $\pm 35$ & $\pm 0.006$ & $\pm 0.06$ & $\pm 0.008$ & $\pm 0.08$ & $\pm 0.025$ & $\pm 0.12$ & $-0.058$ & $+0.059$ & 0.064 \\ 
MB99 & $\pm 35$ & $\pm 0.007$ & $\pm 0.06$ & $\pm 0.009$ & $\pm 0.08$ & $\pm 0.021$ & $\pm 0.11$ & $-0.040$ & $+0.043$ & 0.048 \\ 
\hline
\multicolumn{11}{c}{$\mu$~Leo} \\
VALD3 & $\pm 60$ & $\mp 0.003$ & $\pm 0.09$ & $\pm 0.008$ & $\pm 0.15$ & $\pm 0.052$ & $^{+0.24}_{-0.23}$ & $-0.095$ & $+0.106$ & 0.114 \\ 
MB99 & $\pm 60$ & $\mp 0.004$ & $\pm 0.09$ & $\pm 0.017$ & $\pm 0.15$ & $\pm 0.040$ & $\pm 0.17$ & $-0.052$ & $+0.061$ & 0.071 
 \enddata 
\tablenotetext{}{The $\sigma_p$ and the
 $\Delta (p)$ indicate the error of stellar parameter $p$ and its effect on $\log\epsFe$, where
 $p$ takes $\Teff$, $\log g$, or 
 [M/H]. The
 ${\Delta_{\rm b}}^{\pm}$
 indicate the error of $\log\epsFe$ from the bootstrap
 method. In the last column,
 the ${\Delta_{\rm total}}$ is the total uncertainty (see details in text).}
\end{deluxetable*}

\subsection{Comparison with previous results}
\label{sec:literature} 
\begin{figure}[!tb]
 \begin{center}
 
\includegraphics[clip,width=0.98\hsize]{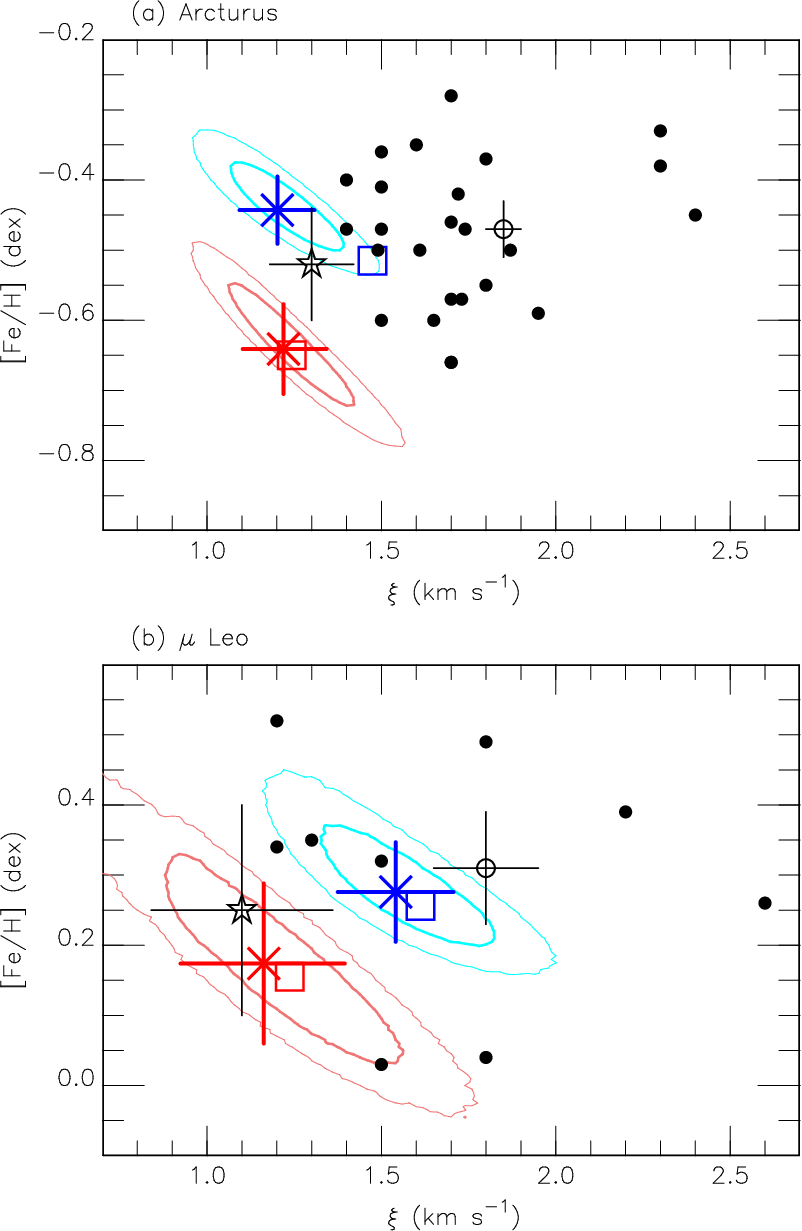} \caption{Comparison of
our estimates of $\xi$ and [Fe/H] with previous estimates. The
contours, cross symbols, and open squares are as in
Figure~\ref{fig4}. The results for VALD3 and MB99 are illustrated in red
and blue, respectively. The total errors in Table~\ref{tab5} are added
to the crosses. Two recent results in the literature are shown with
error bars: \citet{Smith-2013} and \citet{Jofre-2014} indicated by an
open circle and a star symbol, respectively. The filled circles indicate the
other previous estimates that we compiled in Figure~\ref{fig2} except
those without the microturbulence explicitly given. \label{fig8}}
 \end{center}
 \end{figure}

Figure~\ref{fig8} plots the scaled metallicity
$\FeH$, where the solar $\log \epsFe$
is assumed to be 7.45~dex, against $\xi$. We compared our iron abundances with those in previous papers (an open circle:
\citealt{Smith-2013}, a star: \citealt{Jofre-2014}, filled circles: the
others) that we compiled in Figure~\ref{fig2} except those without the microturbulence explicitly given. Our total errors are
comparable with the errors estimated by \citet{Smith-2013} and
\citet{Jofre-2014}. Within the errors and scatters of
 [Fe/H] in
the literature, our metallicities based on the $z^\prime$, $Y$, and $J$
bands spectra agree very well with previous estimates. The metallicities
estimated with MB99 show better agreement with previous estimates than
those with VALD3. Considering also that the scatters in
Figure~\ref{fig5} are smaller with MB99, we believe that the
 $\log gf$ values of MB99 are better than those of VALD3 for chemical
abundance analyses.

\section{Concluding remarks}
We used the $z^\prime YJ$ band high-resolution spectra of
Arcturus and $\mu$~Leo, obtained with WINERED, to estimate the
microturbulence and iron abundance with a precision similar to that of
previous results from spectra at different wavelengths. Our lists of
\ion{Fe}{1} lines in the 0.91--1.33~{$\mu$}m range will be useful for
obtaining the precise metallicities of stars obscured by severe
interstellar extinction compared with the optical regime, for which the
extinction is stronger. For many objects in the Galactic disk found in
recent infrared surveys, this new wavelength window may be ideal for
detailed abundance analyses. One of the major error sources is the
uncertainty in $\xi$ in various studies, including ours, based on
spectra at different wavelengths from the optical \citep[e.g., Table 3 of
][]{Jofre-2014} to the $H$-band \citep[e.g., Table 7 of ][]{Smith-2013}.
Furthermore, how to determine the microturbulence and its error is not
established or straightforward. The bootstrap method that we
demonstrated in this paper can give quantitative estimates of the
microturbulence and its error. The error in microturbulence is
 0.11--0.24 $\kms$ for each combination of target and line
list. The obtained microturbulences are consistent with those that were
estimated or assumed in previous studies on the same targets. Note, however, that using different line lists (or different sets of lines)
can result in slightly different microturbulences depending especially on
the $\log gf$ values of strong lines used in the
analysis. The very strong lines ($X>-6$) were rejected because these lines
are likely to introduce problems into a chemical abundance analyses
due to severe saturation, non-LTE effects, the contribution of EW from
the damping wing, and so on. 
Considering the comparison of our estimates with previous ones in addition to the scatters of $\log
 \epsFe$, we adopt the measurements with the \ion{Fe}{1} lines selected
 from MB99 as our best estimates: $(\xi, \log\epsFe)=(1.20\pm0.11\
 \kms,7.01\pm 0.05~{\rm dex})$ and $(1.54\pm0.17\ \kms,7.73\pm 0.07{\rm
 ~dex})$ for Arcturus and $\mu$~Leo, respectively.


\acknowledgments

We acknowledge useful comments from the anonymous referee.
We are grateful to the staff of Koyama Astronomical Observatory for
their support during our observation. We thank Yoichi Takeda for
providing us with SPTOOL. This work has made use of the VALD database,
operated at Uppsala University, the Institute of Astronomy RAS in
Moscow, and the University of Vienna. This study has been financially
supported by Grants-in-Aid (numbers 16684001, 20340042, 21840052,
26287028, and 18H01248) from the Japan Society for the Promotion of
Science (JSPS) and by Supported Programs for the Strategic Research
Foundation at Private Universities (S0801061 and S1411028) from the
Ministry of Education, Culture, Sports, Science and Technology (MEXT) of
Japan. K.F. is supported by a JSPS Grant-in-Aid for Reseearch Activity
Start-up (No.~16H07323). N.K. is supported by JSPS-DST under the
Japan-India Science Cooperative Programs during 2013-2015 and 2016-2018.

\end{document}